\documentclass[graybox]{svmult}

\usepackage{mathptmx}       
\usepackage{helvet}         
\usepackage{courier}        
\usepackage{type1cm}        

\usepackage{amsmath}
\usepackage{amssymb}

\usepackage{makeidx}         
\usepackage{graphicx}        
\usepackage{multicol}        
\usepackage[bottom]{footmisc}

\def\beq{\begin{eqnarray}}
\def\eeq{\end{eqnarray}}
\def\ln{\,\mbox{ln}\,}

\def\al{\alpha}
\def\be{\beta}

\def\ka{\kappa}
\def\la{\lambda}

\makeindex             

\begin{document}


\title*{Gravitational ``seesaw'' and light bending in higher-derivative gravity}
\author{Antonio Accioly, Breno L. Giacchini and Ilya L. Shapiro}
\institute{Antonio Accioly, Breno L. Giacchini \at Centro Brasileiro de Pesquisas F\'{\i}sicas, Rua Dr. Xavier Sigaud 150, Urca, 22290-180, Rio de Janeiro, RJ, Brazil, \email{accioly@cbpf.br, breno@cbpf.br}
\and Ilya L. Shapiro \at Departamento de F\'{\i}sica, ICE, Universidade Federal de Juiz de Fora, Campus Universit\'{a}rio - Juiz de Fora, 36036-330, MG, Brazil; Tomsk State Pedagogical University and Tomsk State University, Tomsk, Russia, \email{shapiro@fisica.ufjf.br}}

\maketitle

\abstract*{Local gravitational theories with more than four derivatives have
remarkable quantum properties, e.g., they are super-renormalizable and may
be unitary in the Lee-Wick sense. Therefore, it is important to explore also the IR limit of these theories and identify observable signatures of the higher derivatives. In the present work we study the scattering of a photon by a classical external gravitational field in the sixth-derivative model whose propagator contains only real, simple poles. Also, we discuss the possibility of a gravitational seesaw-like mechanism, which could allow the make up of a relatively small physical mass from the huge massive parameters of the action. If possible, this mechanism would be a way out of the Planck suppression, affecting the gravitational deflection of low energy photons. It turns out that the mechanism which actually occurs works only to shift heavier masses to the further UV region. This fact may be favourable for protecting the theory from instabilities, but makes experimental detection of higher derivatives more difficult}

\abstract{Local gravitational theories with more than four derivatives have
remarkable quantum properties, e.g., they are super-renormalizable and may
be unitary in the Lee-Wick sense. Therefore, it is important to explore also the IR limit of these theories and identify observable signatures of the higher derivatives. In the present work we study the scattering of a photon by a classical external gravitational field in the sixth-derivative model whose propagator contains only real, simple poles. Also, we discuss the possibility of a gravitational seesaw-like mechanism, which could allow the make up of a relatively small physical mass from the huge massive parameters of the action. If possible, this mechanism would be a way out of the Planck suppression, affecting the gravitational deflection of low energy photons. It turns out that the mechanism which actually occurs works only to shift heavier masses to the further UV region. This fact may be favourable for protecting the theory from instabilities, but makes experimental detection of higher derivatives more difficult.}

\section{Introduction}
\label{sec:Intro}

The idea of including higher-derivative terms in the Einstein-Hilbert action was proposed still in the early years of general relativity, and was considered more seriously during the 1960's and 1970's driven by quantum theoretical considerations. Indeed, the renormalization of quantum fields on curved space-time requires the introduction of curvature-squared terms~\cite{UtDW}; also, it was shown that the fourth-derivative gravity is renormalizable, in opposition to the Einsteinian theory~\cite{Stelle77}. As it is widely known, this type of theory usually suffer from Ostrogradsky instabilities at the classical level and have negative-norm states when quantized; notwithstanding, in absence of a straight road to quantum gravity, the role played by higher-derivative terms should be investigated. In this spirit, it was recently shown that gravity theories with more than four derivatives are super-renormalizable~\cite{highderi}, and may yield a unitary S-matrix in the Lee-Wick sense if all the massive poles in the propagator are complex~\cite{LQG-D4}. Some other recent studies on general super-renormalizable theories can be found in Refs.~\cite{Newton-high,seesaw,Giacchini,Larger}.

In the present work we study the bending of light in the most simple super-renormalizable gravity theory, i.e., the sixth-derivative model described by the action
\begin{eqnarray}
S
&=&
S_{\text{grav}}\,+\,\int d^4 x \sqrt{-g} \,{\cal L}_\text{m}\,,
\label{totaction}
\\
S_{\text{grav}}
&=&
\int d^4 x \sqrt{-g} \left\lbrace 
\dfrac{2}{\kappa^2} R + \dfrac{\alpha}{2} R^2
+ \dfrac{\beta}{2} R_{\mu\nu}^2 + \dfrac{A}{2} R \square R
+ \dfrac{B}{2} R_{\mu\nu} \square R^{\mu\nu}\right\rbrace  \,,
\label{Lag6orderGravity}
\end{eqnarray}
where an additional matter action was introduced. Here $\al$,
$\be$, $A$ and $B$ are free parameters; the first two are
dimensionless while $A$ and $B$ carry dimension of (mass)$^{-2}$.
The notation $\kappa^2/2=16\pi G = M_P^{-2}$ is conventional in
the quantum gravity literature; here $M_P$ is the Planck mass.

In the section~\ref{sec:bending} we discuss the deflection of light caused by a static massive body within the semi-classical framework, while in section~\ref{sec:seesaw} we analyse the possibility of avoiding Planck suppression effects to this phenomenon due to a specific seesaw-like mechanism. Our conclusions are summarized in the section~\ref{sec:finale}. We note that further consideration on the issues presented in this work can be found in~\cite{seesaw,Larger}.

Our sign convention follows from the definitions $\eta_{\mu\nu} = \text{diag} (1,-1,-1,-1)$, ${R^\rho}_{\lambda\mu\nu} = \partial_\mu {\Gamma^\rho}_{\lambda\nu} + \cdots$ and $R_{\mu\nu} = {R^\rho}_{\mu\nu\rho}$. Also, we set $\hbar = c = 1$.

\section{Light bending in the sixth-order gravity}
\label{sec:bending}

In the weak field regime we consider the metric to be a fluctuation around the flat-space, $g_{\mu\nu} = \eta_{\mu\nu} + \kappa h _{\mu\nu}$, with $\vert \kappa h _{\mu\nu} \vert \ll 1$. Then, it is possible to show that the field generated by a static point-like mass, has non-zero components given by~\cite{Larger}
\begin{eqnarray}
\label{solutionFinal}
h_{00}
&=&
\frac{M\kappa}{16\pi} \Big( - \frac{1}{r}
+ \frac{4}{3} F_2 - \frac{1}{3}F_0\Big)\,,
\nonumber
\\
h_{11}
&=&
 h_{22} \,=\, h_{33} \,=\, \frac{M\kappa}{16\pi}
 \Big( - \frac{1}{r} + \frac{2}{3}\,F_2 +\frac{1}{3}\,F_0\Big)\,,
\end{eqnarray}
where
\begin{equation}
F_k =
\frac{m_{k+}^2}{m_{k+}^2 - m_{k-}^2} \frac{e^{-m_{k-}r}}{r}
+ \frac{m_{k-}^2}{m_{k-}^2 - m_{k+}^2} \frac{e^{-m_{k+}r}}{r}\,.
\nonumber
\end{equation}

Here $k=0,2$ labels the spin of the particles, whose masses are
defined by the positions of the poles of the propagator,
\begin{equation}
\label{Def_masses}
m_{2\pm}^2
= \dfrac{\beta \pm \sqrt{\beta^2 + \frac{16}{\kappa^2}B}}{2 B},
\quad
m_{0\pm}^2 =
\dfrac{\sigma_1 \pm \sqrt{\sigma_1^2
- \frac{8\sigma_2}{\kappa^2}}}{2 \sigma_2},
\quad
\sigma_1 = 3\alpha + \beta,
\,\,\,
\sigma_2 = 3A+B\,.
\end{equation}
As mentioned above, in this work we assume that the parameters $\alpha,\beta,A$ and $B$ are such that $m_{k\pm} \in \mathbb{R}$ and $m_{k+} \neq m_{k-}$ (for the most general scenario see~\cite{Larger}). In particular, it must hold that $\beta, B < 0$. It is possible to show that $m_{2+}$ and  $m_{0+}$ are ghost modes, while the others are healthy excitations~\cite{Newton-high}.

\begin{figure}[b]
\sidecaption
\includegraphics[scale=0.95]{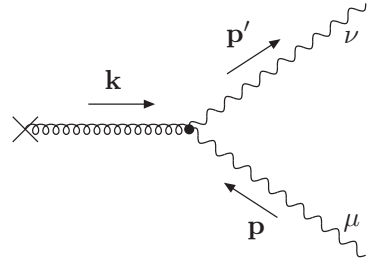}
%
%
\caption{Photon scattering by an external gravitational field.
Here $\vert\textbf{p}\vert \approx \vert\textbf{p}^\prime\vert$.}
\label{Fig1}       
\end{figure}

The deflection of light due to a weak gravitational field can be evaluated within the semi-classical approach by considering the photon to be a quantum particle which interacts with the classical external field~\eqref{solutionFinal}. At tree-level the only diagram contributing to the scattering is the one depicted in Fig.~\ref{Fig1}, which produces the vertex function
\begin{eqnarray}
\label{vertice}
V_{\mu\nu}(p,p^\prime)
&=&
\frac{\ka}{2}\, h^{\lambda\rho}_{\text{ext}} (\textbf{k})
\big[ -\eta_{\mu\nu}\eta_{\lambda\rho}\, p\cdot p^\prime
+ \eta_{\la\rho}p^\prime_\mu p_\nu
\\
\nonumber
&+&
2 \left( \eta_{\mu\nu}p_\lambda p_\rho^\prime - \eta_{\nu\rho}
 p_\lambda p_\mu^\prime - \eta_{\mu\lambda}p_\nu p_\rho^\prime
 + \eta_{\mu\lambda}\eta_{\nu\rho} p\cdot p^\prime \right)\big] .
\end{eqnarray}

Neglecting
energy exchange between the photon and gravitational field and assuming that the bending angle is small, it is possible to show that the unpolarized cross section for this process reads~\cite{seesaw}
\beq
\label{CrossSection}
\dfrac{d\sigma}{d\Omega}
= 16 G^2 M^2
\left[ \frac{1}{\theta^2}
+ \frac{E^2}{m_{2-}^2 - m_{2+}^2}
\Big( \dfrac{m_{2+}^2}{E^2\theta^2 + m_{2-}^2}
- \dfrac{m_{2-}^2}{E^2\theta^2 + m_{2+}^2}\Big)\right] ^2
\eeq
where $E=E^\prime$ is the energy of the photon and $\theta$ is the
deflection angle, i.e., the angle encompassed by $\textbf{p}$
and $\textbf{p}^\prime$.

From the previous expression it is possible to conclude that
\begin{itemize}
\item[i.] light deflection does not depend on $m_{0\pm}$, and thus on the sectors $R^2$ and $R\square R$. This happens because these sectors can be regarded as conformal transformations on the metric~\cite{Larger}.
\item[ii.] Light deflects less than in general relativity. In fact, the ghost $m_{2+}$ gives opposite-sign effect compared to the healthy massive mode  $m_{2-}$ and the graviton, and
\begin{equation}
\label{Eq103}
m_{2-} > m_{2+} \,\,\,\, \Longrightarrow \,\,\,\, \frac{m_{2-}^2 E^2}{E^2\theta^2 + m_{2+}^2}
>
\frac{m_{2+}^2 E^2}{E^2\theta^2 + m_{2-}^2} \,\,\,\, \Longrightarrow \,\,\,\, \left( \frac{d\sigma}{d\Omega} \right)_{E}
>
\frac{d\sigma}{d\Omega}
>
 0 ,
\end{equation}
where $\left( d\sigma / d\Omega \right)_{E} = ( 4 G M / \theta^2 )^2$ is the cross-section for general relativity.
\item[iii.] The scattering is dispersive -- more energetic photons undergo less deflection. In fact, the second relation in~\eqref{Eq103} shows that among the dispersive interactions, the repulsive one is stronger. Therefore, since all the photons are equally attracted by the $R$-sector, the more energetic ones are more repelled and thus less scattered.
\end{itemize}

In order to evaluate the deflection undergone by a photon with energy $E$ and impact parameter $b$ we can compare the previous expression to the classical cross-section formula $d\sigma / d\Omega \, = \, - \, b \, \theta^{-1} db / d\theta$, which yields
\beq
\label{AngleSemiclassical}
\frac{1}{\theta_E^2}
&=&
\frac{1}{\theta^2} + \frac{E^2}{(m_{2-}^2
- m_{2+}^2)^2} \left( \frac{m_{2-}^4}{E^2\theta^2 + m_{2+}^2}
+ \frac{m_{2+}^4}{E^2\theta^2 + m_{2-}^2} \right)
\nonumber
\\
&+&
\frac{2 E^2}{m_{2-}^2 - m_{2+}^2}
\Bigg[ \frac{m_{2-}^2}{m_{2+}^2}\,
\ln\left(
\frac{E^2\theta^2}{E^2\theta^2 + m_{2+}^2} \right)
- \frac{m_{2+}^2}{m_{2-}^2} \,\ln\left(
\frac{E^2\theta^2}{E^2\theta^2+ m_{2-}^2} \right)
\nonumber
\\
&-&
 \frac{m_{2-}^2 m_{2+}^2}{(m_{2-}^2 - m_{2+}^2)^2}\,
\ln\left(\frac{E^2\theta^2 + m_{2-}^2}{E^2\theta^2 + m_{2+}^2}
\right)
\Bigg] ,
\eeq
where $\theta_E = 4 G M / b$ is the scattering angle in Einstein's gravity.

The effect of both massive modes is related to the ratio $E/m_{2\pm}$, in such a manner that photons with transplanckian energies would not be deflected at all, while sufficiently low-energetic photons are scattered just like as in general relativity. Only at an intermediate scale of energy there is a non-trivial scattering.

In particular, it is possible to conceive a scenario in which the hierarchy between the masses is so strong, i.e. $m_{2-} \sim M_P \gg m_{2+}$, that the effect of higher derivatives could be perceived even for the energy scale currently measured, emitted by astrophysical sources. (At the same time, the influence of the healthy massive mode is negligible.) Due to the analogy with the seesaw mechanism of netrino Physics -- in which large-mass parameters combine to yield physical masses with strong hierarchy -- we shall call this possibility as the gravitational seesaw. Under these circumstances, the equation for the deflection angle~\eqref{AngleSemiclassical} reduces to
\beq
\label{AngleSemiclassical_4th}
\frac{1}{\theta_E^2}
= \frac{1}{\theta^2} + \dfrac{E^2}{E^2 \theta^2 + m_{2+}^2}
+ \frac{2E^2}{m_{2+}^2} \ln \dfrac{E^2 \theta^2}{E^2 \theta^2
+ m_{2+}^2} ,
\eeq
which is the same expression that occurs in the fourth-derivative gravity~\cite{Accioly15}.

\section{On the gravitational seesaw}
\label{sec:seesaw}

From Eq.~\eqref{Def_masses} it is straightforward to derive the seesaw condition for the masses $m_{2\pm}$:
\begin{equation}
16 \vert B \vert \ll \kappa^2 \beta^2 .
\end{equation}
If this condition is satisfied, the masses $m_{2\pm}$ can be approximated by
\begin{equation}
m_{2+}^2 \approx \frac{4}{\kappa^2 \vert \beta \vert}
\,\ll\,
m_{2-}^2 \approx \frac{\beta}{B}.
\label{8}
\end{equation}

As in the original neutrino's seesaw mechanism one of the masses
depends, roughly, on only one parameter, while the other depends
on both. There is, however, a remarkable difference with respect to the neutrino's
case: while there it works to make the
lightest mass even lighter, in gravity the effect
is to shift the largest mass further to the UV region, according to Eq.~\eqref{8}. In fact, if the
lighter mass is reduced, then the larger mass is augmented. 
This happens because of the parameter
$B$ which occurs in the denominator of Eq.~\eqref{Def_masses}; indeed, it easy to verify that $m_{2+}$ is a decreasing function on $B$. Thus, the only form of reducing the lightest mass by changing the sixth-derivative parameter is to make it tend to zero (remember that $B<0$); this procedure makes the ghost mode to approach the mass of the fourth-derivative gravity tensor excitation~\cite{Stelle77} as shows Eq.~\eqref{8}. As a consequence, in order to have $m_{2+} \ll m_{2-} \sim M_P$ one must have $\beta \gg 1$.

In this spirit, now focusing our attention on the healthy mode, there are two possible ways of having $m_{2-}$ of the
order of the Planck mass: to have a small $|B|$ or a large $|\beta|$.
The former is the standard choice, since it prescribes that
$\beta \sim 1$ and $B \sim M_P^{-2}$ so as to have all the masses
to the order of $M_P$. The latter relies on the
seesaw mechanism, allowing one to have $|B| \gg M_P^{-2}$ and
still have $m_{2-} \sim M_P$. Of course, having a large $|B|$ still
yielding one large mass can only be achieved on account of the ghost
mass reduction trough a parameter $\beta \gg 1$.

Therefore, the much lighter mass of the first ghost
depends only on the second- and fourth-derivative terms; and the higher-order ones cannot produce an efficient seesaw mechanism working like in the case of the neutrino mass. Only a ``weak seesaw'' is possible, i.e., the reduction of the lightest mass by having a huge dimensionless parameter $\beta$. (See~\cite{Larger} for a discussion of this result in the complex poles case; and~\cite{seesaw} for the generalization to the case of arbitrary-order local models.)

Let us now return to the deflection angle equation~\eqref{AngleSemiclassical_4th} in the presence of the ``weak seesaw''. We notice that the energy of the photon and the quantity $m_{2+}$ always appear through the ratio $m_{2+}/E$. Thus, one can fix the scattering angle at a slightly different figure from general relativity's one -- this could be, e.g., the experimental accuracy of a set of detectors, say $\theta = \theta_E - \Delta \theta$ -- and solve the equation for the aforementioned ratio. For example, if we set $\theta = 1.65" = \theta_E - 0.10"$ for a photon just grazing the Sun, then Eq.~\eqref{AngleSemiclassical_4th} yields
\beq
\frac{m_{2+}^2}{E^2} \,=\, 4.30 \times 10^{-9}\,.
\label{bound}
\eeq

This equation relates the energy of the photon and the mass of the lightest particle necessary to cause a deviation of $0.1"$ from general relativity's prediction. For example, considering that this is the accuracy of the measurements carried out in the visible spectrum during solar eclipses~\cite{Jones,Schmeidler}, it follows the lower bound $m_{2+} \gtrsim 10^{-13}$~GeV. This limit is still very far from the Planck scale, and only with much higher frequencies it is expected that the massive modes could be detected.

\section{Conclusions}
\label{sec:finale}

We have described the bending of light in the sixth-derivative super-renormalizable gravity theory, in the particular case that the propagator has only real, simple poles. Among the main conclusions of this semi-classical analysis we mention the fact that light is less scattered than in general relativity, and that more energetic photons undergo less deflection. A seesaw-like mechanism which could, in principle, avoid the Planck suppression to one of the masses was also proposed. We showed, however, that differently from the neutrino's one, the gravitational seesaw can only work to make the largest mass even larger, on account of the reduction of the smallest one. Therefore, the only possibility of having a small physical mass (while the other is of the order of $M_P$) is to have a huge $\beta$. This makes the experimental detection of higher-derivatives more difficult, but is favourable for protecting the theory from instabilities.

\begin{acknowledgement}
A.A. acknowledges CNPq and FAPERJ for support. 
B.L.G. is thankful to CNPq for supporting his Ph.D. project. 
I.Sh. is grateful to CNPq, FAPEMIG
and ICTP for partial support of his work.
\end{acknowledgement}


\begin{thebibliography}{0}


\bibitem{seesaw} A. Accioly, B.L. Giacchini, I.L. Shapiro,
{\it On the gravitational seesaw and light bending}, arXiv:1604.07348.

\bibitem{Larger} A. Accioly, B.L. Giacchini, I.L. Shapiro,
{\it Low-energy effects in a higher-derivative gravity model with real and complex massive poles}, arXiv:1610.05260.

\bibitem{Accioly15}
A. Accioly, J. Helay\"el-Neto, B. Giacchini, W. Herdy,
 Phys. Rev. D {\bf 91}, 125009 (2015).

\bibitem{highderi} M. Asorey, J.L. L\'opez, I.L. Shapiro,
Int. Journ. Mod. Phys. A {\bf 12}, 5711 (1997).

\bibitem{Giacchini} B.L. Giacchini,
{\it On the cancellation of Newtonian singularities in higher-derivative
gravity}, arXiv:1609.05432.


\bibitem{Jones} B. F. Jones,  	
Astron. J. {\bf 81}, 455 (1976).

\bibitem{Newton-high} L. Modesto, T. de Paula Netto, I.L. Shapiro,
JHEP {\bf 1504}, 098 (2015).

\bibitem{LQG-D4} L. Modesto, I.L. Shapiro,
Phys. Lett. B {\bf 755}, 279 (2016).

\bibitem{Schmeidler} F. Schmeidler,  	
Astron. Nachr. {\bf 306}, 77 (1985).


\bibitem{Stelle77} K. Stelle, Phys. Rev. D {\bf 16}, 953 (1977).


\bibitem{UtDW}
R. Utiyama, B.S. DeWitt, J. Math. Phys. {\bf 3}, 608 (1962).






\end{thebibliography}
\end{document}